\documentclass[12pt]{iopart}

\usepackage[dvips]{graphicx}

\begin{document}

\title[Evolving  small-world networks with geographical attachment preference]{Evolving  small-world networks with geographical attachment preference}

\author{Zhongzhi Zhang}
\address{
Institute of Systems Engineering, Dalian University of Technology,
\\ Dalian 116024, Liaoning, China}%
\ead{dlutzzz063@yahoo.com.cn}
\author{Lili Rong}
\address{
Institute of Systems Engineering, Dalian University of Technology,
\\ Dalian 116024, Liaoning, China}%
\ead{llrong@dlut.edu.cn}
\author{Francesc Comellas}
\address{
Dep. de Matem\`atica Aplicada IV, EPSC, Universitat
Polit\`ecnica de Catalunya\\
 Av. Canal Ol\'{\i}mpic s/n, 08860
Castelldefels, Barcelona, Catalonia, Spain}
\ead{comellas@mat.upc.es}

\begin{abstract}
We introduce a minimal extended evolving model for small-world
networks which is controlled by a parameter. In this
model the network growth is determined by the attachment of new nodes to already
existing nodes that are geographically close. We analyze
several topological properties for our model both analytically and by
numerical simulations. The resulting network shows some important
characteristics of real-life networks such as the small-world
effect and a high clustering.

\end{abstract}



\maketitle


\section{Introduction}
Many real-life systems display both a high degree of local
clustering and the small-world effect~\cite{AlBa02,DoMe02,Ne03,Ne00}.
Local clustering characterizes the tendency of groups of nodes to be
all connected to each other, while the small-world effect describes
the property that any two nodes in the system can be connected by
relatively short paths. Networks with these two characteristics are
called small-world networks.

In the last few years, a number of models have been proposed to
describe real-life systems with small-world effect. The first and
the most widely-studied model is the simple and attractive small-world network
model of Watts and Strogatz (WS model)~\cite{WaSt98}, which
triggered a sharp interest in the studies of the different properties
of small-world networks~\cite{AlBa02,DoMe02,Ne03,Ne00}.
Barth\'el\'emy and Amaral studied the origins of the small-world
behavior in Ref.~\cite{BaAm99}. Barrat and Weigt addressed
analytically as well as numerically the structure properties of the WS
model~\cite{BaWi00}. Amaral \emph{et al.} investigated the
statistical characteristics of a variety of diverse real-life
networks~\cite{AmScBaSt00}. Latora and Marchiori introduced the
concept of efficiency of a network and found that small-world networks
are both globally and locally efficient~\cite{LaMa01}. In
Refs.~\cite{PaAm99,Mo99,PaAm01,NeWa99b}, the spread and percolation
properties were investigated, dealing with the spread of information
and disease along the shortest path in the graph or the spread along
the spanning tree. Recently, researchers have also focused their
attention on other different aspects, characterizing many properties
of small-world networks~\cite{NiMoLaFrHo02,MeHoMiKi03,HuZoTaShJi03,HeSa03,BrBuCoHaSt03,GuKo04,BlKr05}.

In addition to the above-mentioned aspects, variations of the WS
model are another focus of recent interest. Of these variants,
a model proposed independently by Monasson~\cite{Mona99}
and by Newman and Watts~\cite{NeWa99a}, has been thoroughly
studied~\cite{NeWa99b,NeMoWa00}. In 1999, Kasturirangan
presented an alternative version to the WS model~\cite{Ka99}, a special
case of which is exactly solvable~\cite{DoMe00}. One year later,
Kleinberg provided a generalization of the WS model which is based
on a two-dimensional lattice~\cite{Kl00a,Kl00b}. The above
models are all random. In fact, small-world networks can be also
created by deterministic techniques such as modifications of some
regular graphs~\cite{CoOzPe00}, addition and product of
graphs~\cite{CoSa02}. During the past few years, networks generated
in deterministic ways have been also intensively
studied~\cite{BaRaVi01,IgYa05,DoGoMe02,JuKiKa02,RaBa03,No03,
CoFeRa04,ZhWaHuCh04,AnHeAnSi05,DoMa05,ZhCoFeRo05}.

All the above models may partially mimic aspects of real-life small-world
networks. Furthermore, these models are probably
reasonable illustrations of how some networks are shaped. However,
the small-world effect is much more general,
and it is of interest to investigate other mechanisms producing
small-world networks. Recently, Ozik, Hunt and Ott have introduced a
simple evolution model (OHO model) of growing small-world networks
with geographical attachment preference, in which all connections
are made locally to geographically nearby sites~\cite{OzHuOt04}.
Zhang, Rong and Guo have presented a deterministic small-world model
(ZRG model) created by edge iterations~\cite{ZhRoGo05}, which is a
deterministic version of a special case of the OHO model and a variant
of the pseudofractal scale-free network~\cite{DoGoMe02}.

The OHO model and ZRG model may provide valuable insights into some
existing real-world systems. It is then a natural question whether
there is an encompassing scheme, which can put these two specific
models into a more general perspective. In this paper, we propose a
general scenario for constructing evolving small-world networks.
Similar to the OHO and ZRG models, in our model, when a new node
is added to the network, it is only connected to those
preexisting nodes that are
geographically close to it.
Our model results in an exponential degree
distribution, a large clustering coefficient and small average
path length (APL), with values close to those known for many random
small-world networks~\cite{WaSt98,NeWa99a,NeWa99b,Ka99,
DoMe00,Kl00a,Kl00b,OzHuOt04}.
Interestingly, our model includes a parameter $q$ which
controls part of the structural properties of the evolving
small-world networks. Moreover, by tuning this parameter, one can
obtain the OHO model and the ZRG model as particular cases of our model.

The rest of this paper is organized as follows. Section 2 provides a
detailed description of the construction for this evolving
small-world network model. In Section 3, we give  analytical and
simulation results of the main network properties: Degree distribution,
clustering coefficient and average path length. The final section provides
some conclusions.

\section{Evolving small-world network model}
In this section we describe a model of growing network, which is
constructed in an iterative manner. We denote our network after $t$
time steps by $N(t)$. Then the network is constructed in the
following way. We start from an initial state ($t=0$) of
$m+1$ ($m$ even) nodes distributed on a ring all
of which are connected to one another. For $t\geq 1$, $N(t)$ is
obtained from $N(t-1)$ as follows: For each internode interval along the
ring of $N(t-1)$, with probability $q$, a new node is created and
connected its $m$ nearest neighbors
($\frac{m}{2}$ on either side) previously existing at step $t-1$.
Distance, in this case, refers to the number of intervals
along the ring. The growing process is
repeated until the network reaches the desired size. Figure~1 shows
the network growing process for a special case of $m=2$ and $q=1$.

\begin{figure}
\begin{center}
\includegraphics[width=12cm]{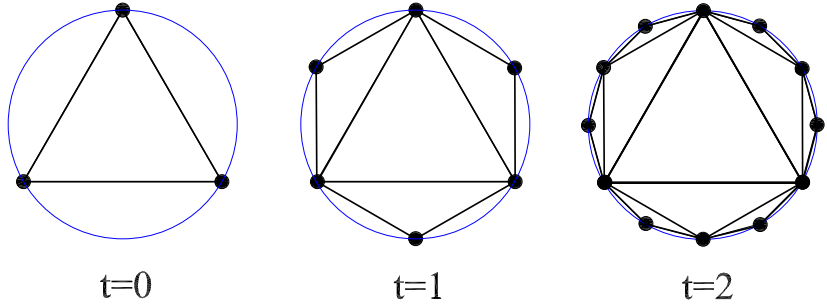}
\caption{Illustration of the growing small-world network for $m=2$
and $q=1$, showing the first three steps of the iterative process. }
\label{netfig}
\end{center}
\end{figure}

When $q=1$ and $m=2$, the network is reduced to the deterministic
ZRG model~\cite{ZhRoGo05}. If $q<1$, the network is growing
randomly. Especially, as $q$ approaches to zero (without reaching
this value) the model coincides with the OHO model~\cite{OzHuOt04},
where at each time step, only one interval is chosen and linked to
its $m$ nearest neighbors, with every interval having the same
probability of being selected (see~\cite{Do03} for interpretation).
Varying $q$ in the interval (0,1) allows one to study the crossover
between the OHO model~\cite{OzHuOt04} and the ZRG
model~\cite{ZhRoGo05}. It should be mentioned that as $q$ is a real
number, below we will assume that all variables concerned with $q$
change continuously. Notice that similar presumption has been used
in Refs.~\cite{AlBa02,DoMe02,Ne03}, which is valid in the limit of
large $t$.

Now we compute the number of nodes and edges of $N(t)$.
We denote the number of newly added nodes and edges at step $t$ by
$L_v(t)$ and $L_e(t)$,
respectively. Thus, initially ($t=0$), we have $L_v(0)=m+1$ nodes
and $L_e(0)=m(m+1)/2$ edges in $N(0)$. Let $N_c (t)$ denote the
total number of internode intervals along the ring
at step $t$, then $N_c (0)= m+1$. By construction, we have
$L_v(t)= N_c (t-1)q$
for arbitrary $t \geq 1$. Note that, when a new node
is added to the network, an interval is destroyed and replaced by
two new intervals, hence the number of total intervals increases by
one. Thus, we have the following relation: $N_c(t)= N_c(t-1)+L_v(t)$.
On the other hand, the addition of each new node leads to $m$ new
edges, after simple calculations one can obtain that at $ t_i $
($t_i \geq 1$), $L_v(t_i)=(m+1)(1+q)^{t_i-1}q$ and
$L_e(t_i)=m(m+1)(1+q)^{t_i-1}q$, respectively. Therefore, the number of nodes
$N_t$ and the total of edges $E_t$ of $N(t)$ is
\begin{eqnarray}\label{Nt}
N_t=\sum_{t_j=0}^{t}L_v(t_j)=(m+1)(1+q)^{t}
\end{eqnarray} and
\begin{eqnarray}\label{Et}
E_t=\sum_{t_j=0}^{t}L_e(t_j)=m(m+1)\left[(1+q)^{t}-\frac{1}{2}\right]
\end{eqnarray}
respectively. The average node degree is then
\begin{equation}
<k>_{t}=\frac{2E_t}{N_t}=2m\left[1-\frac{1}{2(1+q)^{t}}\right ]
\end{equation}
For large $t$ and any $q\neq 0$, it is small and approximately equal
to $2m$. Notice that many real-life networks are sparse in the sense
that the number of edges in the network is much less than
$N_{t}(N_{t}-1)/2$, the number of all possible
edges~\cite{AlBa02,DoMe02,Ne03}.

\section{Structural properties of the evolving small-world Network}
Structural properties of the networks are of fundamental
significance to understand the complex dynamics of real-life
systems. Here we focus on four important characteristics: degree
distribution, clustering coefficient, average path length and
diameter.
\subsection{Degree distribution}
Degree is the simplest and most intensively studied
characteristic of an individual node. The degree of a node $i$ is
the number of edges in the whole network connected to $i$. The
degree distribution $P(k)$ is defined as the probability that a randomly
selected node has exactly $k$ edges. Let $k_{i}(t)$ denote the
degree of node $i$ at step $t$. If node $i$ is added to the
network at step $t_i$ then, by construction, $k_{i}( t_i)=m$. In each
of the subsequent time steps, there are $m$ intervals with
$\frac{m}{2}$ at each side of $i$.
Each of these intervals could be considered, with  probability $q$, to
create a new node connected to $i$. Then
the degree $k_i(t)$ of node $i$ satisfies the relation
\begin{equation}
k_{i}(t)=k_{i}(t-1)+mq
\end{equation}
considering the initial condition $k_{i}( t_i)=m$, we obtain
\begin{equation}\label{Ki}
k_{i}(t)=m+mq(t-t_{i})
\end{equation}
The degree of each node can be obtained explicitly as in
Eq.~\ref{Ki}, and we see that this degree increases at each
iteration. So it is convenient to obtain the cumulative
distribution~\cite{Ne03}
\begin{equation} \label{cumulative distribution1}
P_{cum}(k)=\sum_{k'=k}^{\infty}P(k')
\end{equation}
which is the probability that the degree is greater than or equal to
$k$. An important advantage of the cumulative distribution is that it can
reduce the noise in the tail of probability distribution. Moreover,
for some networks whose degree distributions have exponential tails:
$P(\tilde{k}) \sim e^{-\tilde{k}/\kappa}$, cumulative distribution
also gives exponential expression with the same exponent:
\begin{equation} \label{cumulative distribution2}
P_{cum}(\tilde{k})=\sum_{k'=\tilde{k}}^{\infty}P(k')\sim
\sum_{k'=\tilde{k}}^{\infty}e^{-k'/\kappa}\sim e^{-\tilde{k}/\kappa}
\end{equation}
This makes exponential distributions particularly easy to spot
experimentally, by plotting the corresponding cumulative
distributions on semilogarithmic scales.

Using Eq.~\ref{Ki}, we have $P_{cum}(k)=\sum_{k'=k}^{\infty}P(k)=
P\left (t'\leq\tau=t-(\frac{k-m}{mq})\right)$. Hence
\begin{eqnarray}\label{cumulative distribution3}
P_{cum}(k)=\sum_{t'=0}^{\tau}\frac{L_v(t')}{N_{t}}
=\frac{m+1}{(m+1)(1+q)^{t}}+\sum_{t'=1}^{\tau}\frac{(m+1)(1+q)^{t'-1}q}{(m+1)(1+q)^{t}}\nonumber
\\=(1+q)^{-\frac{k-m}{mq}}
\end{eqnarray}
The cumulative distribution decays exponentially with $k$. Thus the
resulting network is an exponential network. Note that most
small-world networks including the WS model belong to this
class~\cite{BaWi00}.

\begin{figure}
\begin{center}
\includegraphics[width=9cm]{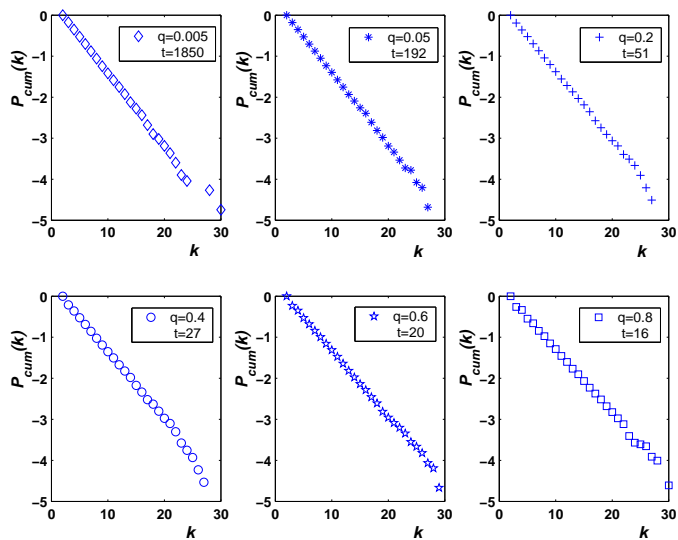}
\caption{Semilogarithmic graph of the cumulative degree distribution for
the evolving networks in the case of  $m=2$ and for different values of $q$. All
data points are obtained by averaging ten independent simulations.} \label{degree}
\end{center}
\end{figure}

In Fig.~(\ref{degree}), we report the simulation results of
the cumulative degree distribution for several values of $q$ and with
$m=2$. Except in the deterministic case $q=1$, the degree spectrum
of the networks is continuous. From Fig.~(\ref{degree}), we can see
that the cumulative degree distribution decays exponentially for
large degree values, in agreement with the analytical results and
supporting a relatively homogeneous topology similar to most small-world
networks~\cite{WaSt98,NeWa99a,NeWa99b,Ka99,DoMe00,Kl00a,Kl00b,OzHuOt04}.
Other values of $m$ should give qualitatively a similar behavior as for $m=2$.

\subsection{Clustering coefficient}
Most real-life networks show a cluster structure which can be quantified
by the clustering coefficient~\cite{AlBa02,DoMe02,Ne03,Ne00}.
The clustering of a node gives the relation of connections of the
neighborhood nodes closest to it.
By definition, the clustering of
a node $i$ with $k_{i}$ adjacent nodes is given by
$C_{i}=2e_{i}/[k_{i}(k_{i}-1)]$, where $e_{i}$ is the number of
existing edges between its neighbors. The clustering coefficient $C$
of a network is obtained by averaging $C_{i}$ over all the vertices
in the network.

For the particular case  $m=2$, using the connection rules, it is
straightforward to calculate exactly the clustering coefficient of an
arbitrary node and the average value for the network. When a
node $i$ enters the network,  $k_{i}$ and $e_{i}$ are $2$
and $1$, respectively. After that, if the degree $k_{i}$ increases
by one, then its new neighbor must connect one of its existing
neighbors, i.e. $e_{i}$ increases by one at the same time.
Therefore, $e_{i}$ is equal to $k_{i}-1$ for all vertices at all
time steps. So there exists a one-to-one correspondence between the
degree of a node and its clustering. For a node $v$ with degree
$k$, the exact expression for its clustering coefficient is $2/k$,
which has been also been obtained in
Ref.~\cite{DoGoMe02,ZhRoGo05,DoGoMe01}. This expression for the
local clustering shows the same inverse proportionality with the
degree than the observed in a variety of real-life
networks~\cite{RaBa03}.

\begin{figure}
\begin{center}
\includegraphics[width=8cm]{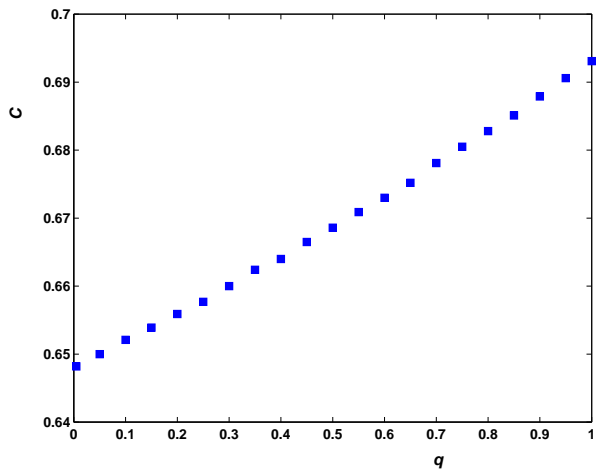}
\caption{Average clustering coefficient $C$ vs $q$ when
$m=2$. Each data point is an average over ten independent
simulation runs.}
\label{clustering}
\end{center}
\end{figure}

In addition to the good scaling of the clustering coefficient for single
node, the average clustering coefficient $C$ of the network is
very high. Also, $C$ depends on $q$ and approaches to a constant
asymptotic value as the network order is very large. In
Fig.~(\ref{clustering}), we show $C$ as a function of $q$ in the case
of $m=2$. From Fig.~(\ref{clustering}), one can see in the infinite
order limit of the network, that $C$ approaches to a nonzero constant value.
Simulations exhibit that $C$ equals to $0.6482$, $0.6560$, $0.6640$,
$0.6729$ and $0.6828$ for $q=0.005$, $0.2$, $0.4$, $0.6$, and $0.8$,
respectively. Fig.~(\ref{clustering}) reflects the dependence of  $C$, the
clustering coefficient of the network, on $q$. It is obvious that
$C$ increases continuously with $q$. As $q$ increases from 0 to 1,
$C$ grows from $\frac{3}{2}\ln 3-1$~\cite{OzHuOt04} to
$\ln 2$~\cite{ZhRoGo05}, i.e. from $0.6479$ to $0.6931$. The reason for
this dependence relation would need further study, but might be related
to a biased choice of the edges chosen at each iteration, see Ref.~\cite{CoRobA05}. Although we only focus on the case  $m=2$, one expects that for other values of $m$,
$C$ also will converge to a different nonzero value for every different value of $q$
(see Ref.~\cite{OzHuOt04} for a particular case).

\subsection{Average path length}
Certainly, the most important property for an small-world network is
a logarithmic average path length (APL) (with the number of nodes).
It has obvious implications for the dynamics of processes taking
place on networks. Therefore, its study has attracted much attention.
Here APL means the minimum number of edges connecting a pair of
nodes, averaged over all pairs of nodes.
Below, using an approach similar to that presented
in~\cite{ZhYaWa05}, we will study the APL of our network for the
particular case $m=2$.

We label each of the network nodes according to their creation
times, $v=1,2,3,\ldots,N-1,N.$ We denote $L(N)$ as the APL of our
network with order $N$. It follows that
$L(N)=\frac{2\varepsilon(N)}{N(N-1)}$, where $\varepsilon(N)=\sum_{1
\leq i<j \leq N}\ell_{i,j}$ is the total distance, where
$\ell_{i,j}$ is the smallest distance between node $i$ and $j$.

\begin{figure}
\begin{center}
\includegraphics[width=8cm]{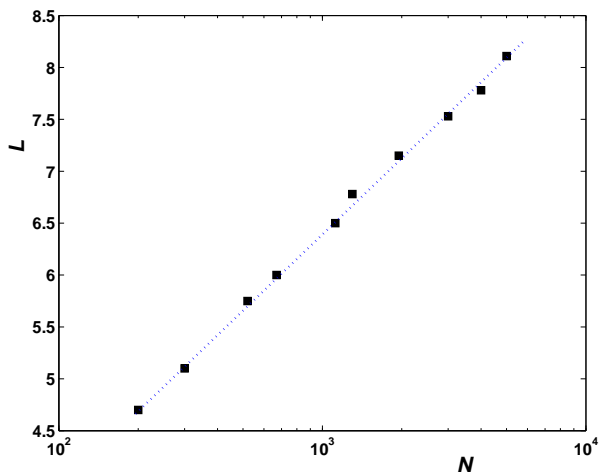}
\caption{Semilogarithmic graph of the dependence of average path
length on network order $N$ in the case of $m=2$ and $q=0.5$. All
values plotted are averages over ten independent realizations. The
values can be fitted well by a straight line.} \label{distance}
\end{center}
\end{figure}

For this special case  $m=2$, any newly-created node is actually
only attached to both ends of an edge. Thus the distances
between existing node pairs will not be affected by the addition
of new vertices. Then we have the following equation:
\begin{equation}\label{Eq8}
L(N+1)=L(N)+ \sum_{i=1}^{N}\ell_{i,N+1}
\end{equation}
Like in the analysis of~\cite{ZhYaWa05,ZhRoCo05}, Eq.~(\ref{Eq8}) can be
rewritten approximately as:
\begin{equation}\label{Eq10}
L(N+1) \approx L(N)+N+(N-2)L(N-1)
\end{equation}
After some derivations, we can provide an upper bound for the
variation of $\varepsilon(N)$ as
\begin{equation}
{d\varepsilon(N) \over dN} =  N + {2\varepsilon(N) \over N}
\end{equation}
which leads to
\begin{equation}\label{Eq17}
\varepsilon(N) = N^2\ln N + \beta,
\end{equation}
where $\beta$ is a constant. As $\varepsilon(N) \sim N^2\ln N $, we
have $L(N) \sim \ln N$. Therefore, we have proved that in the
special case of $m=2$ of our model, there is an slow growth
of the APL with the network size $N$. In Fig.~(\ref{distance}), we present
the APL vs the network order $N$ in the case of $m=2$ and $q=0.5$. We
see that the APL behaves logarithmically as a function of $N$. We
expect that for other values of $q$, the APL will present a similar behavior. In
fact, in the case of $q=1$, we can compute exactly the diameter of
the network (i.e. the maximum distance between all pairs of nodes).
A sharp analytical proof shows that the diameter also grows
logarithmically with the number of nodes of the network~\cite{ZhRoGo05}. It should
be noted that in our model, considering values of $m$ greater than 2,
then the APL will increase more slowly than in the case $m=2$ as in those cases the
larger $m$ is, the denser the network becomes.

Similar to Refs.~\cite{OzHuOt04,ZhRoGo05}, the interpretation for
the slow growth of APL is as follows. The older nodes that had once
been geographically proximal along the ring are pushed apart as new
nodes are positioned in the interval between them. From
Fig.~\ref{netfig} we can see that when new nodes enter into the
network, the original nodes are not near but, rather, have many
newer nodes inserted between them. Thus, the network growth creates
"shortcuts" attached to old nodes, which join remote nodes along the
ring one another as in the WS model~\cite{WaSt98}.

\subsection{Diameter for deterministic networks}
As we have mentioned above the diameter of a network is the maximum of
the distances between all pairs of nodes, characterizing the longest
communication delay in the network. Small diameter is consistent
with the concept of small-world. In the deterministic case  $q=1$,
we denote $N(t)$ as $N_{q=1}(t)$ and $Diam(N_{q=1}(t))$ as the
diameter of $N_{q=1}(t)$ which can be computed exactly. But here we
only give an upper bound on the diameter. The obtained bound scales
logarithmically with the order of the networks. Now we present the
main ideas of this analysis as follows.

Clearly, at step $t=0$, $Diam(N_{q=1}(0))$ equals to 1. At each step
$t\geq 1$, we call newly-created nodes at this step \emph{active
nodes}. Since all active nodes are attached to those nodes existing
in $N_{q=1}(t-1)$, so one can easily see that the maximum distance
between arbitrary active node and those nodes in $N_{q=1}(t-1)$ is
not more than $Diam(N_{q=1}(t-1))+1$ and that the maximum distance
between any pair of active nodes is at most $Diam(N_{q=1}(t-1))+2$.
Thus, at any step, the diameter of the network increases by 2 at
most. Then we get $2(t+1)$ as an upper bound of $Diam(N(t)$. Note
that the logarithm of $N_{q=1}(t)$ is $\ln((m+1) 2^{t})=
t\ln2+\ln(m+1)$, which is approximately equal to $(t+1)\ln 2$ in the
limit of large $t$. Thus the diameter grows at most logarithmically
with the network order. Since our aim here is to show that the
network diameter is small, so we only give a rough upper on diameter
not more exact than that in~\cite{ZhRoGo05}.

\section{Conclusion}
To sum up, we give here a simple evolving model for small-world
networks. During the network growth, new nodes do not have
a complete knowledge of all the current network nodes, but are
attached to those preexisting sites that are geographically close to
them. We have obtained both analytically and numerically the
solution for relevant parameters of the network and we have verified that
our model exhibits
the classical characteristics of small-world network: a high
clustering and a short APL. In addition, the model under consideration
is actually a tunable generalization  which includes as particular
extreme cases the models introduced in
Refs.~\cite{OzHuOt04} and~\cite{ZhRoGo05}. Moreover, the networks
can model a
variety of real-life networks whose topologies are influenced by
such geographical constraints.

\subsection*{Acknowledgment}
This research was supported by the Natural Science Foundation of
China (Grant No. 70431001). Support for F.C. was provided by the
Secretaria de Estado de Universidades e Investigaci\'on (Ministerio
de Educaci\'on y Ciencia),  Spain, and the European Regional
Development Fund (ERDF) under project TIC2002-00155.


\end{document}